\documentclass[aps,prc,groupedaddress,showpacs,manuscript]{revtex4}

\usepackage{amsmath}
\usepackage{graphicx}

\begin{document}

\title{Production and rescattering of strange baryons at SPS energies in a transport model with hadron potentials}

\author {Qingfeng Li,$\, ^{1}$\footnote{E-mail address: liqf@hutc.zj.cn}
and Zhuxia Li$\, ^{2}$ \email[]{lizwux@ciae.ac.cn}}
\address{
1) School of Science, Huzhou Teachers College, Huzhou 313000,
P.R.\ China  \\
2) China Institute of Atomic Energy, P.O.\ Box 275 (18),
Beijing 102413, P.R.\ China\\
 \\
 }


\begin{abstract}
A mean-field potential version of the Ultra-relativistic Quantum
Molecular Dynamics (UrQMD) model is used to investigate the
production of strange baryons, especially the $\Lambda$s and
$\overline{\Lambda}$s, from heavy ion collisions at SPS
energies. It is found that, with the consideration of both formed
and pre-formed hadron potentials in UrQMD, the transverse mass and
longitudinal rapidity distributions of experimental data of both
$\Lambda$s and $\overline{\Lambda}$s can be quantitatively explained
fairly well. Our investigation also shows that both the production
mechanism and the rescattering process of hadrons play important roles
in the final yield of strange baryons.
\end{abstract}


 \maketitle

The heavy ion collision (HIC) is the only way of the human being
at present to explore the properties of nuclear matter at
supranormal and subnormal densities and/or high temperatures. The
properties of the sub-structure and the dynamics of the
nucleus/nucleon could be even discovered by high-energy HICs which
has being explored experimentally by the SchwerIonen Synchrotron
(SIS) at Gesellschaft f\"ur Schwerionenforschung (GSI, Germany),
the Alternating Gradient Synchrotron (AGS) at the Brookhaven
National Laboratory (BNL,USA), the Super Proton Synchrotron (SPS)
at the European Organization for Nuclear Research (CERN,
Switzerland), the Relativistic Heavy Ion Collider (RHIC) at BNL,
and the Large Hadron Collider (LHC) at CERN. Theoretically, the
most important property of the nuclear matter relates closely to
its equation of state (EoS)
\cite{Stoecker:1986ci,Aichelin:1991xy,Cassing:1999es,Baran:2004ih}.
The stiffness of the EoS at both low and high densities have also
being received much attention in past decades
\cite{Stoecker:1981iw,Ko:1983zp,Bombaci:1991zz,Brown:2000pd,Danielewicz:2002pu,Li:2005kqa,Tsang:2008fd}.
At high densities, the stiffness of the EoS will definitely decide
the order and the level of the phase transition to the quark-gluon
plasma (QGP)
\cite{Rischke:1995mt,Spieles:1997ab,Bluhm:2007nu,Li:2008qm,Cassing:2010ge}.
It is a consensus that any final conclusion of it should only be
drawn by large numbers of comparison works between  theoretical
simulations and experimental measurements.

In the past, several signals - such as charmonium suppression,
relative strangeness enhancement, energy loss of hard partons, etc -
of the (phase) transition to the deconfined phase have been observed
in HICs at SPS energies
\cite{Matsui:1986dk,Soff:1999et,Dumitru:2001xa,Wang:2001cs,Heinz:2006ur,Torrieri:2007qy}.
However, none of them gives $100\%$ undoubt conclusion of the
probable phase transition. Especially, the strangeness production
and enhancement is still a matter of argument. It is quite necessary
to investigate the strange hadrons with a microscopic transport
model which might give deeper insights into the whole process of the
production and the transport of all hadrons. The Ultra-relativistic
Quantum Molecular Dynamics (UrQMD)
\cite{Bass98,Bleicher99,Bratkovskaya:2004kv,Zhu:2005qa,Li:2008qm} is
one of the most suitable microscopic transport models which have
been worked fairly successfully in this field for more than 10 years
in addition to the Relativistic Quantum Molecular Dynamics (RQMD)
\cite{Sorge:1989dy}, the Parton-Hadron-String Dynamics (PHSD)
\cite{Cassing:2007nb} and A MultiPhase Transport model (AMPT)
\cite{Lin:2002gc} models.

However, based on previous comparison of UrQMD calculations with
data, we know that there are some disagreement about the strange
hadron production at AGS and SPS energies
\cite{Bass98,LI:2005zi,Petersen:2008kb}. As for hyperons, first, the
yields of hyperons are somewhat overestimated in UrQMD cascade
calculations using versions earlier than $2.1$. Starting from
version 2.1, the UrQMD group considers additional high mass
resonances which leads to a smaller yield of hyperons so that a nice
agreement with data from central Pb+Pb collisions at SPS energies
was seen in previous calculations \cite{Bratkovskaya:2004kv}.
Second, the resulting mean transverse momenta of hyperons were found
to be too low as compared with experimental data when using the
version earlier than $2.1$ \cite{Petersen:2008kb}. It is interesting
to see that this problem is also partly cured by the newest version
2.3. Third, even though the former two problems have been cured
partly with the updating cascade version, the yield of anti-hyperons
decreases at the same time which makes the comparison with data
become even worse \cite{Petersen:2008kb}. These problems attract our
attention and will be discussed in this paper with a mean-field
potential version of the UrQMD model.

In this paper, after a brief introduction of the UrQMD and its
recent updates, the production and the evolution of strange hadrons
as well as anti-protons at SPS energies are investigated with the
mean-field potential version. For comparison, the corresponding
cascade calculation results are also shown. The experimental data
are taken from NA49 collaboration
\cite{Mitrovski:2006js,Blume:2007kw,Anticic:2009ie}. Finally, a
conclusion and outlook is given.

The UrQMD model is a microscopic transport approach based on the
covariant propagation of constituent (anti-)quarks and diquarks
accompanied by mesons and baryons, as well as the corresponding
anti-particles, i.e., full baryon-anti-baryon symmetry is included.
It simulates multiple interactions of ingoing and newly produced
particles, the excitation and fragmentation of color strings  and
the formation and decay of hadronic resonances
\cite{Bass98,Bleicher99}. Besides the cascade mode in which all
particles are treated to be free streaming between collisions, it is
also necessary to incorporate the mean-field contribution for a
complete dynamic transport \cite{Li:2005gfa,Li:2007yd,Li:2008qm}.
Since the UrQMD model inherits the basic treatment of the baryonic
equation of motion in the QMD model \cite{Aichelin:1986wa}, the
consideration of the mean-field contribution is also logical. In
order to properly describe the physical process in HICs at SIS
energies this term has been treated carefully in the UrQMD model
before \cite{Petersen:2006vm,Trautmann:2009kq,Yuan:2010ad}. With the
increase of beam energy from SIS, AGS, SPS, up to RHIC, the dynamics
of the transport has being attracted more and more attention due to
the fact that quite a few of discrepancies between cascade
calculations, of any microscopic transport models, and experimental
data have been shown in these beam energy regions, such as the
collective flows \cite{Bleicher:2000sx,Petersen:2006vm}, the nuclear
stopping \cite{Bratkovskaya:2004kv,Yuan:2010ad}, and the HBT
two-particle correlation \cite{Lisa:2005dd,Li:2007yd}.

Recent update of the UrQMD model follows three different routines.
The first one is called as ``cascade version'' which is to modify
the cascade process of UrQMD. This is the main routine and the
newest official version is v2.3 \cite{Petersen:2008kb}. The second
one is called as ``mean-field potential version'' which further
considers the mean-field potentials of both formed and pre-formed
hadrons \cite{Li:2007yd}. The third one is called as ``hybrid
version'' with which the microscopic transport process is
incorporated with a macroscopic hydrodynamics
\cite{Petersen:2008dd,Li:2008qm}. So, by using the 'hybrid version'
one can compare calculations with various EoS during the
hydrodynamic evolution and those with the pure cascade calculations
within the same framework. The first routine is of importance since
it supplies a better basis as a cascade process. The latter two
versions consider more deeply about the dynamic process of all
particles but with different strategies: The second routine bases on
the same structure of the equation of motion at low SIS beam
energies, in which both the mean field and the collision terms
should be taken into account explicitly. While the third routine
replaces the UrQMD dynamic process in between the initialization and
the hadron rescattering processes by a hydrodynamic one. Therefore,
one sees clearly that, with the rapid development of researches on
the dynamic process of high-energy HICs, the modifications on the
transport model are still quite frequent and effective.

Starting from the version 2.0, similar to HSD, the PYTHIA
\cite{Sjostrand:2006za} is considered  in order to treat the
initial hard collisions more carefully, which is important for
HICs at high SPS and RHIC energies. In the version 2.1
\cite{Bratkovskaya:2004kv}, similar to RQMD, the high-mass
resonances are re-treated in order to give higher
mean transverse momenta of most particles. The newest version 2.3
\cite{Petersen:2008kb} is then brought out with some other minor
changes.  The mean-field potential version is based on the cascade
version v2.1 but not v2.3 which is partly due to the minor
difference between them. As for the mean-field contribution, in
addition to the conventional potentials for formed
hadrons \cite{Li:2005gfa}, the mean-field potentials for
pre-formed baryons from string fragmentation have also been taken
into account \cite{Li:2007yd}, i.e., the Yukawa, the Coulomb, and
the momentum dependent terms are neglected but the similar density
dependent (Skyrme-like) term as the formed baryons is used, which
reads as

\begin{equation}
U(\rho_h/\rho_0)=a (\frac{\rho_h}{\rho_0})+b
(\frac{\rho_h}{\rho_0})^g, \label{den1}
\end{equation} while a reduction factor (2/3) is considered for
pre-formed mesons due to the quark-number
difference. In Eq.\ (\ref{den1}) $\rho_0=0.16 fm^{-3}$ is the
normal nuclear density. $a$, $b$, and $g$ are parameters, in this
work for the SM-EoS, they are $-110$ MeV, $182$ MeV, and $7/6$,
respectively. The $\rho_h$ is the hadronic density, which reads as

\begin{equation} \rho_h=\sum_{j\neq i}c_i c_j\rho_{ij}
\label{den2}\end{equation} where $c_{i,j}=1$ for baryons, $2/3$ for
pre-formed mesons, and $0$ for formed mesons. $\rho_{ij}$ is a
Gaussian in coordinate space.

As in Ref.\ \cite{Isse:2005nk}, the relativistic effects on the
relative distance ${\bf r}_{ij}={\bf r}_i-{\bf r}_j$ and the
relative momentum ${\bf p}_{ij}={\bf p}_i-{\bf p}_j$ employed in the
two-body potentials (Lorentz transformation) are considered as
follows:

\begin{equation}
{\bf \tilde{r}_{ij}^2}={\bf r}_{ij}^2+\gamma_{ij}^2({\bf
r}_{ij}\cdot {\bf \beta}_{ij})^2, \label{rboost}
\end{equation}
\begin{equation}
{\bf \tilde{p}_{ij}^2}={\bf
p}_{ij}^2-(E_i-E_j)^2+\gamma_{ij}^2(\frac{m_i^2-m_j^2}{E_i+E_j})^2.
\label{pboost}
\end{equation}
In Eqs.~\ref{rboost} and \ref{pboost} the velocity factor ${\bf
\beta}_{ij}$ and the corresponding $\gamma$-factor of $i$ and $j$
particles are defined as $ {\bf \beta}_{ij}=({\bf p}_i+{\bf
p}_j)/(E_i+E_j)$ and $ \gamma_{ij}=1/\sqrt{1-{\bf \beta}_{ij}^2}$.
Furthermore, a covariance-related reduction factor for potentials in
the Hamiltonian, $m_j/E_j$, was introduced in the simplified version
of RQMD model \cite{Maruyama:1996rn} and adopted in this work as
well.

\begin{figure}
\includegraphics[angle=0,width=0.8\textwidth]{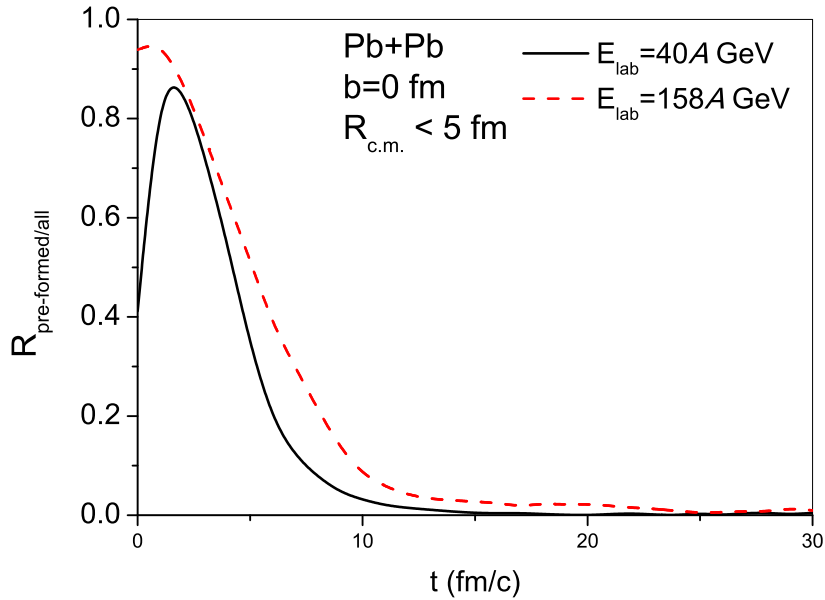}
\caption{Time evolution of the ratio of pre-formed hadrons and all
hadrons (``$R_{pre-formed/all}$'') in the zone $R_{c.m.} < 5 fm$
from central Pb+Pb reaction at $40A$ GeV (solid line) and $158A$ GeV
(dashed line), respectively.} \label{fig1}
\end{figure}

We have found from the HBT correlations of two-particles at AGS,
SPS, and RHIC energies that the mean-field potentials of both formed
and pre-formed hadrons are essential for explaining the sources of
the HBT time-related puzzle \cite{Li:2007yd,Li:2008ge}. Furthermore,
by using the hybrid model, it is found that the equation of state at
high densities should be somewhat stiff in order to explain the HBT
data at SPS energies \cite{Li:2008qm}. Although, the stiffness of
the EoS is still with some uncertainties, it is no doubt that the
dynamic process deserves more investigations. We notice that the
average hadronic density at a central zone (e.g., $R_{c.m.} < 5 fm$)
from central Pb+Pb reactions at SPS energies reaches 4-8 times
normal nuclear density at the early stage of the collision. And, it
is known that in the same beam energy region and at the early stage,
the string process starts to replace the resonance-decay process for
producing new particles in the UrQMD model description. Fig.\
\ref{fig1} shows the time evolution of the ratio of pre-formed
hadrons and all hadrons (``$R_{pre-formed/all}$'') in the zone
$R_{c.m.} < 5 fm$ from the central Pb+Pb reaction at $40A$ GeV
(solid line) and $158A$ GeV (dashed line), respectively. It is seen
clearly that the main production mechanism at $t<5$ fm$/c$ is the
string excitation and fragmentation. And it is up to $t\sim 10$ fm
$/c$ that this production mechanism still plays visible role. Hence,
the pre-formed hadron potentials will definitely provide a large
contribution to the early pressure. Accordingly, the rescattering of
new produced particles will be influenced to some extent, which will
be checked by the current investigation.

\begin{figure}
\includegraphics[angle=0,width=0.8\textwidth]{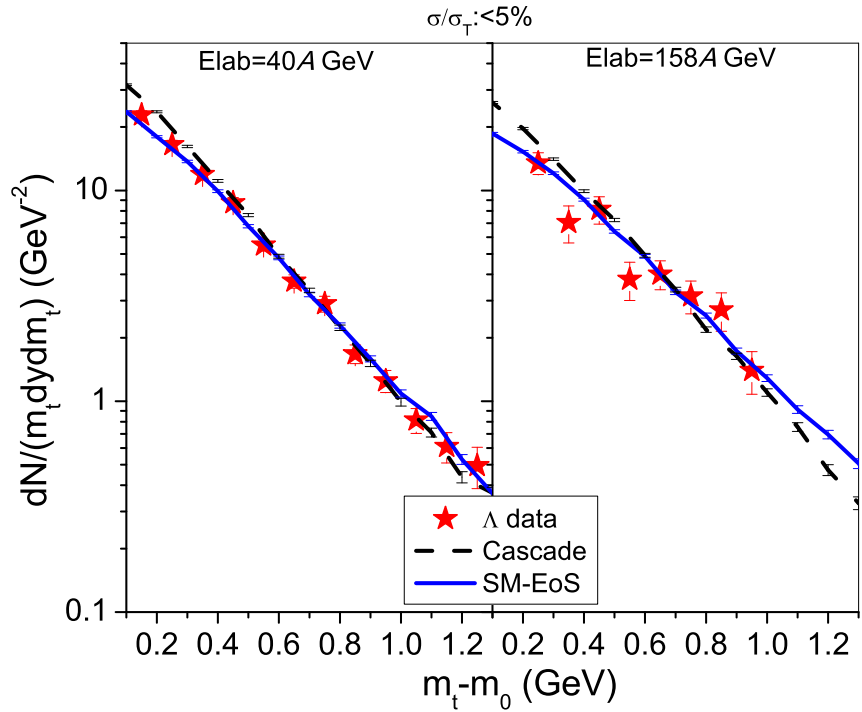}
\caption{Transverse mass spectra of $\Lambda$s ($=\Lambda +
\Sigma^0$) at mid-rapidity ($|y|<0.4$) for central Pb+Pb reactions
($\sigma/\sigma_T<5\%$) at $40A$ GeV (left plot) and at $158A$ GeV
(right plot). The experimental data are shown by stars (taken from
Ref.~\cite{Anticic:2009ie}) while calculations with mean-field
potentials (dashed lines) are compared with those with a cascade
mode (solid lines).} \label{fig2}
\end{figure}

Fig.\ \ref{fig2} shows the transverse mass ($m_t-m_0$, where
$m_t=\sqrt{p_t^2+m_0^2}$ and $m_0$ is the mass of the particle at
rest, $p_t=\sqrt{p_x^2+p_y^2}$ is the transverse momentum of the
particle in the center-of-mass system) spectra of $\Lambda$s
($=\Lambda + \Sigma^0$, same as experimental data) at mid-rapidity
($|y|<0.4$) for central Pb+Pb reactions ($\sigma/\sigma_T<5\%$) at
$40A$ GeV (left plot) and at $158A$ GeV (right plot). The data are
taken from Ref.\ \cite{Anticic:2009ie} for comparison. We find that the
cascade calculation of the UrQMD version $2.1$ already reproduces
the data well within error bars except those at very low and high
transverse masses. We also noticed that if the newest version $2.3$
is used this situation becomes even worse at large transverse
masses, which might be due to the minor changes of the double
strange diquark suppression factor and the single strange diquark
suppression factor. If we use the older version $1.3$, as we have
known before \cite{Petersen:2008kb}, the slope of the transverse
mass spectra is steeper than the newest one. Hence the careful
treatment of the high-mass resonances is very important. With the
consideration of both formed and pre-formed hadron mean-field
potentials, it is found that the slope becomes more flat so that one
can further describe the data at both the low and the high
transverse masses. Therefore, the treatment of the mean field
contributions is also of importance.

\begin{figure}
\includegraphics[angle=0,width=0.8\textwidth]{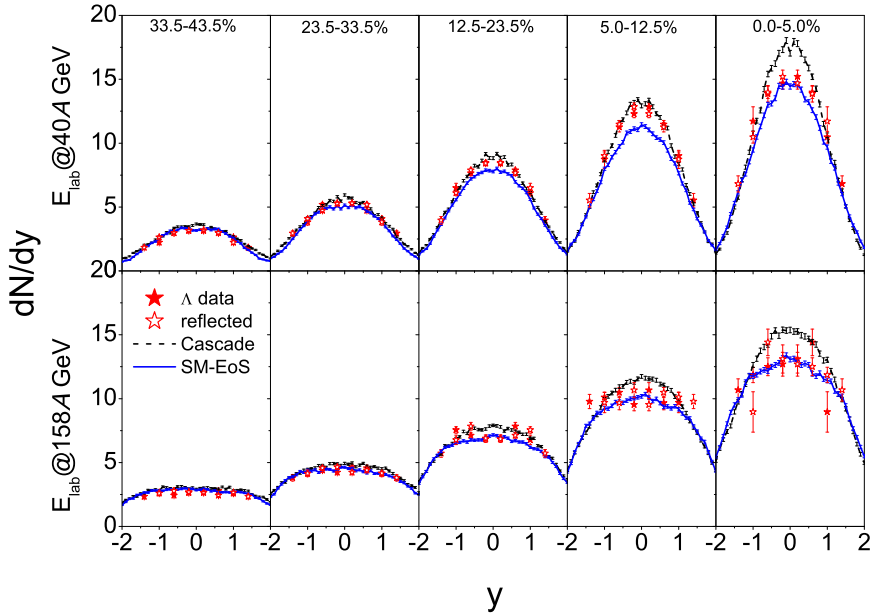}
\caption{Rapidity spectra of $\Lambda$s for Pb+Pb collisions at
$40A$ GeV (upper plots) and $158A$ GeV (lower plots) for 5 different
centrality bins which are shown from the right to the left. The experimental data in each plot is shown by stars and taken from \cite{Anticic:2009ie}.
Calculations with mean-field potentials and with a pure cascade mode
are shown by lines.} \label{fig3}
\end{figure}

Fig.\ \ref{fig3} depicts centrality dependence of the rapidity
distribution of $\Lambda$ yields for Pb+Pb collisions at $40A$ GeV
(upper plots) and $158A$ GeV (lower plots). Results within five
centrality bins from central ($\sigma/\sigma_T<5\%$) to
semi-peripheral ($33.5\%<\sigma/\sigma_T<43.5\%$) collisions are
shown from the right to the left. First of all, it is seen that
the effect of potentials is stronger in more central collisions,
which is clear. Second, the potential effect is mainly to suppress
the yield of hyperons at mid-rapidity which was also seen in
previous calculations at lower beam energies \cite{LI:2005zi}. It
implies again that the dynamic transport of the hyperon after its
production is important. Third, the suppression effect on the
yield of hyperons provides us with a better fit to data
especially at the most central collisions. We also find that
calculations with the version $2.3$ also present similar
results to the ones with potentials. Finally, it is also seen that
in the most central Pb+Pb collisions at $158A$ GeV, the rapidity
distribution of $\Lambda$ yields calculated with potentials is
slightly expanded. This phenomenon should be related to the
stronger early pressure introduced by the pre-formed hadron
potentials and will be discussed later-on.

\begin{figure}
\includegraphics[angle=0,width=0.8\textwidth]{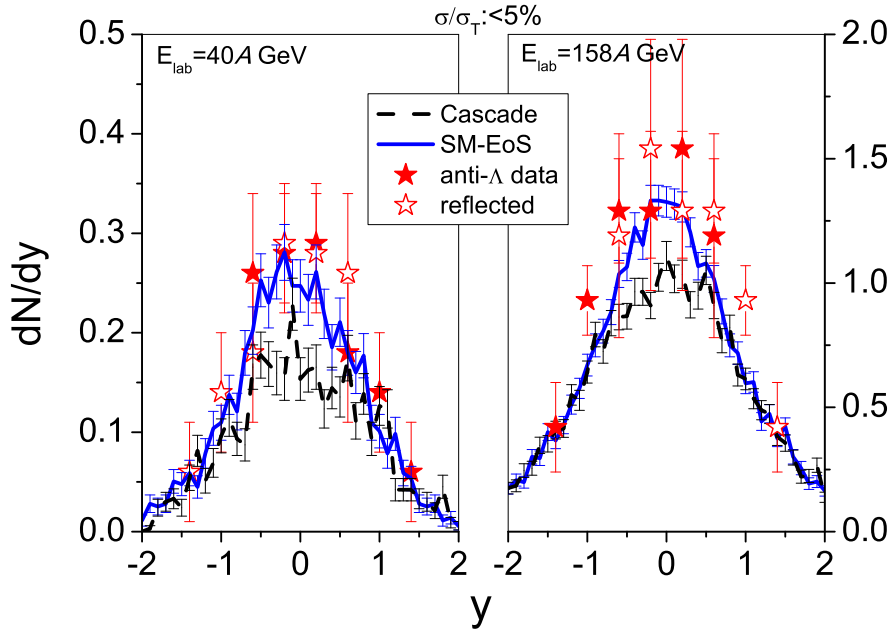}
\caption{Rapidity spectra of $\overline{\Lambda}$s for central Pb+Pb
collisions at $40A$ GeV (left) and $158A$ GeV (right). Experimental
data and the data reflected at mid-rapidity are shown by solid and
open stars \cite{Anticic:2009ie}. Calculations with and without the
mean-field potential are shown by lines. } \label{fig4}
\end{figure}

Fig.\ \ref{fig4} shows the rapidity spectra of $\overline{\Lambda}$s
($=\overline{\Lambda} + \overline{\Sigma^0}$) for central Pb+Pb
collisions at $40A$ GeV (left) and $158A$ GeV (right). Calculations
with and without mean-field hadron potentials are compared to the
data taken from \cite{Anticic:2009ie}. Contrary to the rapidity
distribution of $\Lambda$s, it is interesting to see that the yield
of $\overline{\Lambda}$s at mid-rapidity is driven up to fit the
data quite well when the potentials are considered in calculations.

\begin{figure}
\includegraphics[angle=0,width=0.8\textwidth]{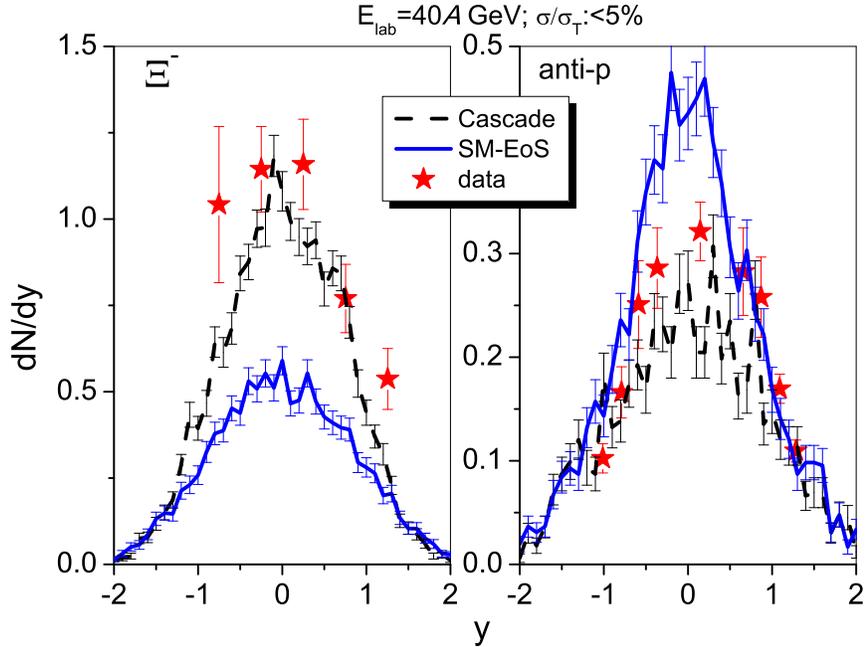}
\caption{Rapidity spectra of $\Xi^-$s (left) and anti-protons
(right) for central Pb+Pb collisions at $40A$ GeV. Experimental data
are shown by stars
\cite{Mitrovski:2006js,Blume:2007kw,Petersen:2008kb}. Calculations
with and without the mean-field potential are shown by lines. }
\label{fig5}
\end{figure}

In order to understand the nice fitting results of both $\Lambda$
and $\overline{\Lambda}$ spectra in both longitudinal rapidity
spectra and transverse mass spectra, and to check the fitting
results of other particles, we show in Fig.\ \ref{fig5} the
rapidity spectra of $\Xi^-$s (left) and anti-protons (right). The
central Pb+Pb collision at $40A$ GeV is chosen as an example since
the experimental data are available for both particles. It is seen
clearly from Fig.\ \ref{fig5} that the calculated $\Xi^-$ yield is
suppressed with the consideration of potentials, which is the same
as $\Lambda$s. While the calculated yield of anti-protons is
enhanced which is the same as anti-$\Lambda$s. At $158A$ GeV, the same
potential effect is seen for both particles.

\begin{figure}
\includegraphics[angle=0,width=0.8\textwidth]{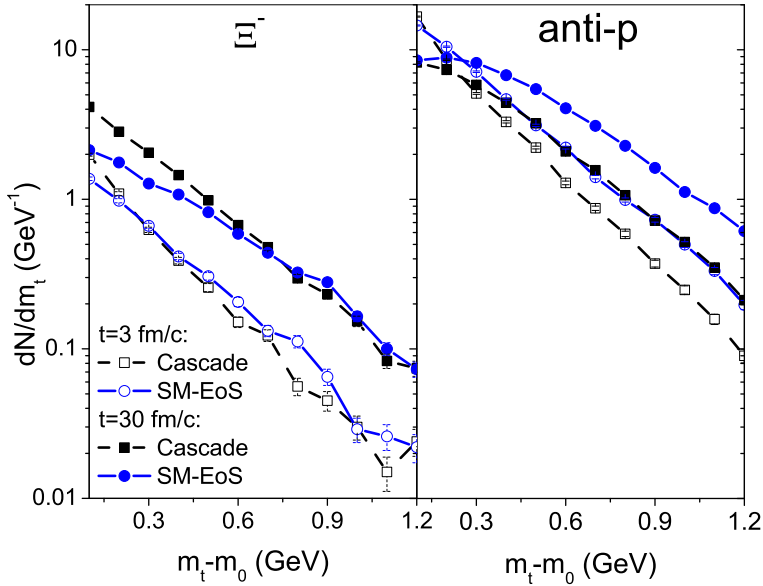}
\caption{Calculated transverse mass spectra of $\Xi^-$s (left) and
anti-protons (right) for central Pb+Pb collisions at $158A$ GeV and
at two time points: $3$ fm$/c$ and $30$ fm$/c$. Calculation results
with and without the mean-field potential are shown by lines with
symbols.} \label{fig6}
\end{figure}

\begin{figure}
\includegraphics[angle=0,width=0.8\textwidth]{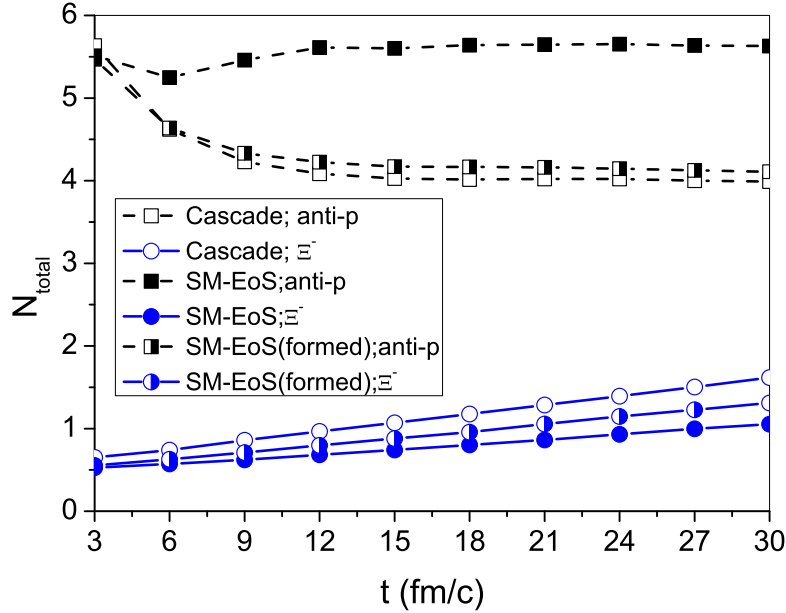}
\caption{Time evolution of total yields of $\Xi^-$s and anti-protons
for central Pb+Pb collisions at $158A$ GeV. Cascade calculation
results (line with open symbols) are compared to the results with
potentials of both formed and pre-formed particles (lines with solid
symbols), and to the results with potentials of only formed
particles (lines with half-open symbols). } \label{fig7}
\end{figure}

Therefore, it is necessary to check the time evolution of these
produced particles in order to understand the potential effects on
both hyperons and anti-particles. Fig.\ \ref{fig6} exhibits the
calculated transverse mass spectra of $\Xi^-$s (left) and
anti-protons (right) at two evolution time points $t=3$ fm$/c$ and
$30$ fm$/c$. The central Pb+Pb reaction at the beam energy $158A$
GeV is chosen in order to have more particles during the time
evolution. From Fig.\ \ref{fig6}, we find that shortly after the
time when the two Lorentz-contracted nuclei have passed through each
other \cite{Steinheimer:2007iy} (e.g., $\sim 3$ fm$/c$ and $\sim
1.5$ fm$/c$ for the Pb+Pb system at $40A$ GeV and $158A$ GeV,
respectively), here we set $t=3$ fm$/c$, the potential effect on
$\Xi^-$ production can be seen at high transverse mass, but it is
more obvious in the anti-proton spectra. It should be due to the
fact that anti-particles are mainly produced at earlier collisions
within a more dense region. Therefore, a higher early pressure
introduced by the pre-formed hadron potential leads to stronger
emission of anti-protons with high momenta. With the increasing time
from $t=3$ fm$/c$ to $t=30$ fm$/c$ (when most of rescatterings of
particles have ceased), we find that in the cascade mode there are
still a large amount of new $\Xi^-$s produced, while the production
of anti-protons in this period becomes much less important. The time
evolution of total yields of particles can be seen more clearly from
Fig.\ \ref{fig7}: in the cascade mode and with the time increasing
from $3$ fm$/c$ to $30$ fm$/c$, the total yield of $\Xi^-$ is
doubled increased, while the yield of anti-protons is about $30\%$
decreased due to the well-known strong annihilation effect. It is
also found that after $12$ fm$/c$ the yield of anti-protons is
almost saturated. With the consideration of formed and pre-formed
hadron potentials, however, more anti-protons are survived which are
mainly due to their higher momenta obtained at the early stage of
reactions. If we switch off the pre-formed hadron potentials but
keep the formed ones, it is seen from Fig.\ \ref{fig7} that the time
evolution of anti-protons is almost the same as that with the
cascade mode. It implies that the mechanism of string excitation and
fragmentation is essential to the production of anti-protons, which
certainly is reasonable. For strange particle production, in
addition to the string mechanism, the rescattering process of
hadrons are also of importance, which can be understood clearly from
the following features shown in Figs.\ \ref{fig6} and \ref{fig7}:
(1) the rapid increase of the $\Xi^-$ yield during the time $3-30$
fm$/c$, (2) the suppression effect of potentials on both the yield
mainly at the low transverse masses shown in Fig.\ \ref{fig6} and
the total yield shown in Fig.\ \ref{fig7} at $t=30$ fm$/c$, and (3)
the contribution of formed hadron potentials to $\Xi^-$ yield (the
line with half-open circle in Fig.\ \ref{fig7}).

Thus, up to now, we have understood more deeply the reason
why the mean-field potentials contribute to the nice fitting
results of both $\Lambda$ and $\overline{\Lambda}$ spectra shown
in Figs.\ \ref{fig2}-\ref{fig4}. However, it is also found from
Figs.\ \ref{fig5}-\ref{fig7} that the modification of the UrQMD
model should carry on with more endeavors. First, the production
of multi-strange baryons is still not satisfactory. The
yield at mid-rapidity is underestimated at SPS energies in all
versions of UrQMD calculations (including the newest version
$2.3$). This problem can not be solved solely by the change of the
double strange diquark suppression factor and/or the single
strange diquark suppression factor. Since the strange-hadron
related cross sections are normally large, the medium effect of
cross sections on the production and the rescattering of
multi-strange baryons should show importance and deserves more
investigations. Second, the production of (anti-) protons at AGS
and SPS energies are still an open problem
\cite{Petersen:2008kb,Yuan:2010ad}. Although the net-proton yield
and the rapidity distributions of (anti-)protons are better
described \cite{Li:2007yd,Yuan:2010ad} by the mean-field potential
version, a further refinement is still required. The dynamic
process of particles in the dense medium induced by both the new
(QGP) phase and the hadronic phase afterwards should be
investigated with a theoretical breakthrough.

To summarize, the production of strange baryons, especially the
$\Lambda$s and $\overline{\Lambda}$s, are investigated with the
mean-field potential version of the UrQMD model for HICs at SPS
energies. In this version of UrQMD, in addition to the
formed hadrons, the mean-field potentials of pre-formed hadrons
are also considered which are similar to those of formed hadrons.
It is found that with the consideration of potentials, the
transverse mass and longitudinal rapidity distributions of
experimental data of both $\Lambda$s and $\overline{\Lambda}$s can
be quantitatively explained fairly well. The early strongly
repulsive and the later-on attractive forces introduced mainly by
the pre-formed and the formed potentials respectively lead to more
$\overline{\Lambda}$ emission but less $\Lambda$ emission at
freeze-out. Our investigation also shows that the hadronic
rescattering process is still important for HICs at SPS energies.

\section*{Acknowledgements}
We would like to thank C. Blume for sending us the experimental data
and M. Bleicher for useful discussions. We acknowledge support by
the computing server C3S2 in Huzhou Teachers College. The work is
supported in part by the key project of the Ministry of Education of
China (No. 209053), the National Natural Science Foundation of China
(Nos. 10905021,10979023), the Zhejiang Provincial Natural Science
Foundation of China (No. Y6090210), and the Qian-Jiang Talents
Project of Zhejiang Province (No. 2010R10102).



\begin{thebibliography}{99}

\bibitem{Stoecker:1986ci}
  H.~Stoecker and W.~Greiner,
  Phys.\ Rept.\  {\bf 137}, 277 (1986).


\bibitem{Aichelin:1991xy}
  J.~Aichelin,
  Phys.\ Rept.\  {\bf 202}, 233 (1991).

\bibitem{Cassing:1999es}
  W.~Cassing and E.~L.~Bratkovskaya,
  Phys.\ Rept.\  {\bf 308}, 65 (1999).

\bibitem{Baran:2004ih}
  V.~Baran, M.~Colonna, V.~Greco and M.~Di Toro,
  Phys.\ Rept.\  {\bf 410}, 335 (2005).

\bibitem{Stoecker:1981iw}
  H.~Stoecker, M.~Gyulassy and J.~Boguta,
  Phys.\ Lett.\  B {\bf 103}, 269 (1981).

\bibitem{Ko:1983zp}
  C.~Ko,
  Phys.\ Lett.\  B {\bf 120}, 294 (1983).

\bibitem{Bombaci:1991zz}
  I.~Bombaci and U.~Lombardo,
  Phys.\ Rev.\  C {\bf 44}, 1892 (1991).

\bibitem{Brown:2000pd}
  B.~A.~Brown,
  Phys.\ Rev.\ Lett.\  {\bf 85}, 5296 (2000).

\bibitem{Danielewicz:2002pu}
  P.~Danielewicz, R.~Lacey and W.~G.~Lynch,
  Science {\bf 298}, 1592 (2002)


\bibitem{Li:2005kqa}
  Q.~Li, Z.~Li, S.~Soff, M.~Bleicher and H.~Stoecker,
  Phys.\ Rev.\  C {\bf 72}, 034613 (2005)

\bibitem{Tsang:2008fd}
  M.~B.~Tsang, Y.~Zhang, P.~Danielewicz, M.~Famiano, Z.~Li, W.~G.~Lynch and A.~W.~Steiner,
  Phys.\ Rev.\ Lett.\  {\bf 102}, 122701 (2009).

\bibitem{Rischke:1995mt}
  D.~H.~Rischke, Y.~Pursun and J.~A.~Maruhn,
  Nucl.\ Phys.\  A {\bf 595} (1995) 383
  [Erratum-ibid.\  A {\bf 596} (1996) 717].

\bibitem{Spieles:1997ab}
 C.~Spieles, H.~St\"ocker and C.~Greiner,
 Phys.\ Rev.\  C {\bf 57} (1998) 908.

\bibitem{Bluhm:2007nu}
  M.~Bluhm, B.~Kampfer, R.~Schulze, D.~Seipt and U.~Heinz,
  Phys.\ Rev.\  C {\bf 76} (2007) 034901.

\bibitem{Li:2008qm}
  Q.~Li, J.~Steinheimer, H.~Petersen, M.~Bleicher and H.~Stocker,
  Phys.\ Lett.\  B {\bf 674}, 111 (2009).

\bibitem{Cassing:2010ge}
  W.~Cassing and E.~L.~Bratkovskaya,
  arXiv:1004.3064 [nucl-th].


\bibitem{Matsui:1986dk}
  T.~Matsui and H.~Satz,
  Phys.\ Lett.\  B {\bf 178}, 416 (1986).


\bibitem{Soff:1999et}
  S.~Soff, S.~A.~Bass, M.~Bleicher, L.~Bravina, E.~Zabrodin, H.~St\"ocker and W.~Greiner,
  Phys.\ Lett.\  B {\bf 471}, 89 (1999).

\bibitem{Dumitru:2001xa}
  A.~Dumitru and R.~D.~Pisarski,
  Phys.\ Lett.\  B {\bf 525}, 95 (2002).

\bibitem{Wang:2001cs}
  E.~Wang and X.~N.~Wang,
  Phys.\ Rev.\ Lett.\  {\bf 87}, 142301 (2001).


\bibitem{Heinz:2006ur}
  U.~Heinz and G.~Kestin,
  PoS C {\bf POD2006}, 038 (2006).


\bibitem{Torrieri:2007qy}
  G.~Torrieri,
  Phys.\ Rev.\  C {\bf 76}, 024903 (2007).


\bibitem{Bass98}S. A. Bass {\it et al.}, [UrQMD-Collaboration], Prog. Part. Nucl. Phys. {\bf 41}, 255 (1998).

\bibitem{Bleicher99}M. Bleicher {\it et al.}, [UrQMD-Collaboration], J. Phys. G: Nucl.
Part. Phys. {\bf 25}, 1859 (1999).

\bibitem{Bratkovskaya:2004kv}
  E.~L.~Bratkovskaya {\it et al.},
  Phys.\ Rev.\ C {\bf 69}, 054907 (2004).

\bibitem{Zhu:2005qa}
  X.~Zhu, M.~Bleicher and H.~Stoecker,
  Phys.\ Rev.\  C {\bf 72}, 064911 (2005).

\bibitem{Sorge:1989dy}
  H.~Sorge, H.~Stoecker and W.~Greiner,
  Annals Phys.\  {\bf 192}, 266 (1989).


\bibitem{Cassing:2007nb}
  W.~Cassing,
  Nucl.\ Phys.\  A {\bf 795}, 70 (2007).

\bibitem{Lin:2002gc}
  Z.~W.~Lin, C.~M.~Ko and S.~Pal,
  Phys.\ Rev.\ Lett.\  {\bf 89}, 152301 (2002).


\bibitem{LI:2005zi}
  Q.~Li, Z.~Li, E.~Zhao and R.~K.~Gupta,
  Phys.\ Rev.\  C {\bf 71}, 054907 (2005).

\bibitem{Petersen:2008kb}
  H.~Petersen, M.~Bleicher, S.~A.~Bass and H.~Stocker,
  arXiv:0805.0567 [hep-ph].


\bibitem{Mitrovski:2006js}
  M.~K.~Mitrovski {\it et al.}  [NA49 Collaboration],
  J.\ Phys.\ G {\bf 32}, S43 (2006).

\bibitem{Blume:2007kw}
  C.~Blume  [Na49 Collaboration],
  J.\ Phys.\ G {\bf 34}, S951 (2007).


\bibitem{Anticic:2009ie}
  T.~Anticic {\it et al.}  [NA49 Collaboration],
  Phys.\ Rev.\  C {\bf 80}, 034906 (2009).

\bibitem{Li:2005gfa}
  Q.~Li, Z.~Li, S.~Soff, M.~Bleicher and H.~Stoecker,
  J.\ Phys.\ G {\bf 32}, 151 (2006).


\bibitem{Li:2007yd}
  Q.~Li, M.~Bleicher and H.~Stocker,
  Phys.\ Lett.\  B {\bf 659}, 525 (2008).

\bibitem{Aichelin:1986wa}
  J.~Aichelin and H.~Stoecker,
  Phys.\ Lett.\  B {\bf 176}, 14 (1986).

\bibitem{Petersen:2006vm}
  H.~Petersen, Q.~Li, X.~Zhu and M.~Bleicher,
  Phys.\ Rev.\  C {\bf 74}, 064908 (2006).

\bibitem{Trautmann:2009kq}
  W.~Trautmann {\it et al.},
  Prog.\ Part.\ Nucl.\ Phys.\  {\bf 62}, 425 (2009).


\bibitem{Yuan:2010ad}
  Y.~Yuan, Q.~Li, Z.~Li and F.~H.~Liu,
  Phys.\ Rev.\  C {\bf 81}, 034913 (2010)
  [Erratum-ibid.\  C {\bf 81}, 069901 (2010)].


\bibitem{Bleicher:2000sx}
  M.~Bleicher and H.~Stoecker,
  Phys.\ Lett.\  B {\bf 526} (2002) 309.

\bibitem{Lisa:2005dd}
  M.~A.~Lisa, S.~Pratt, R.~Soltz and U.~Wiedemann,
  Ann.\ Rev.\ Nucl.\ Part.\ Sci.\  {\bf 55} (2005) 357.

\bibitem{Petersen:2008dd}
  H.~Petersen, J.~Steinheimer, G.~Burau, M.~Bleicher and H.~St\"ocker,
  Phys.\ Rev.\  C {\bf 78} (2008) 044901.

\bibitem{Sjostrand:2006za}
  T.~Sjostrand, S.~Mrenna and P.~Z.~Skands,
  JHEP {\bf 0605}, 026 (2006).

\bibitem{Isse:2005nk}
  M.~Isse, A.~Ohnishi, N.~Otuka, P.~K.~Sahu and Y.~Nara,
  Phys.\ Rev.\  C {\bf 72}, 064908 (2005).

\bibitem{Maruyama:1996rn}
  T.~Maruyama, K.~Niita, T.~Maruyama, S.~Chiba, Y.~Nakahara and A.~Iwamoto,
  Prog.\ Theor.\ Phys.\  {\bf 96}, 263 (1996).

\bibitem{Li:2008ge}
  Q.~Li, M.~Bleicher and H.~Stocker,
  Phys.\ Lett.\  B {\bf 663}, 395 (2008).

\bibitem{Steinheimer:2007iy}
  J.~Steinheimer, M.~Bleicher, H.~Petersen, S.~Schramm, H.~Stocker and D.~Zschiesche,
  Phys.\ Rev.\  C {\bf 77}, 034901 (2008).


\end{thebibliography}
\end{document}